# Title:  A Population Model for the Academic Ecosystem

**Authors:**  Yan Wu[1], Srinivasan Venkatramanan[2], Dah Ming Chiu[1]*

**Affiliations:**

[1] Department of Information Engineering, The Chinese University of Hong Kong, Shatin, N. T. , Hong Kong.

[2] Virginia Bioinformatics Institute, Virginia Tech., Blacksburg, VA, USA.

*Correspondence to: dmchiu@ie.cuhk.edu.hk

**Abstract**: In recent times, the academic ecosystem has seen a tremendous growth in number of authors and publications. While most temporal studies in this area focus on evolution of co-author and citation network structure, this systemic inflation has received very little attention. In this paper, we address this issue by proposing a population model for academia, derived from publication records in the Computer Science domain. We use a generalized branching process as an overarching framework, which enables us to describe the evolution and composition of the research community in a systematic manner. Further, the observed patterns allow us to shed light on researchers' lifecycle encompassing arrival, academic life expectancy, activity, productivity and offspring distribution in the ecosystem. We believe such a study will help develop better bibliometric indices which account for the inflation, and also provide insights into sustainable and efficient resource management for academia.

**Main Text:**

**Introduction:**

Over the past couple of centuries, academia has evolved from a collection of geographically and culturally separated pockets of knowledge production to a thriving global community. A well-established university system and a plethora of public/private funded research institutions have enabled the growth and sustenance of this ecosystem. Researchers communicate their findings by publishing in journals and conferences which can then be consumed by a wider audience, leading to collaborations and citations, thus enabling a faster diffusion of innovative and breakthrough ideas. This has been further accelerated by the presence of Internet, which has brought about sea changes in the way we discover, consume, create and distribute research. It is imperative to understand the causes and consequences of such rapid growth, in order to ensure sustainable and efficient operation of the ecosystem. For instance, by characterizing the growth, we can address issues related to resource bottlenecks (such as the overproduction of PhDs (*1*)), and, in general, efficient resource allocation in the system. Also, with a clear understanding of the inflation (*2*), we can design better ranking mechanisms for authors and papers, taking into account the relative volume of publications generated over the years.

The research community, especially over the past few decades, has seen a tremendous growth in numbers (*2-4*). Considering the field of Computer Science as an example, we observe exponential growth not only in the total number of authors in the system (referred to as *alive authors*) but also in the number of contributing authors (referred to as *active authors*), the number of authors publishing for the first time (referred to as *newcomers*) and in the number of publications each year (Fig. S1 and Fig. 1A).  Such a rapid growth was first quantitatively reported by Price (*5*), who studied research records from 1907 to 1960, and observed an annual growth rate of 4.7% (doubling every 15 years) . Similar observations have been made, in the Physics field (*6*) and the general SCI and SSCI literature databases (*7*). This could be due to two major reasons: (a) authors *merging* from various research communities into the field (referred to as *immigrant*) or (b) sustained production of *offspring* within the field. Further, once they arrive into the mainstream, authors survive for varying amounts of *lifetime*, and through their *activity* and *productivity* lead to future generations of offspring. Previous work studying the academic ecosystem focus more on the structural evolution of coauthor network (*8-11*), with little or no emphasis on the growing size of the system over the years. In this work, taking a cue from demographics and population studies (*12*), we aim to characterize how authors *arrive*, *live*, *stay active*, *produce*, and *reproduce* in the ecosystem, capturing the lifecycle of authors. We use branching processes as an overarching framework to describe this evolution.

**Branching processes:**

Branching processes (BP) (*13, 14*) were first proposed by Galton and Watson to study the extinction of family surnames. They have been subsequently applied to model population growths, in the context of demographics, epidemiology, biochemistry, etc. Several variants have been proposed in the past which consider age-dependence (*15-17*), deaths (*18*), immigration (*13*) and environmental effects (*14*). Since the study of academic population can be likened to demographics, it is plausible to look at it from the BP framework. Academic genealogy projects such as The Mathematics Genealogy Project (*19*) have gathered such data, representing known advisor-advisee pairs as parent-offspring in the BP. The inference of such advisor-advisee links from the publications data was also considered in (*20*). Recently, the BP hypothesis has been used in the academic context, for evaluating the correlations between mentor and protégé performance (*21*).

However, it is necessary to note the key differences between the traditional BP assumptions and the dynamics in academic community. (a) In traditional BP, a parent produces all its offspring *at the same time* and consequently, the reproduction phases of the parent and offspring are *temporally separated*. However in academia, a parent can continue producing offspring until it leaves the system, thus overlapping with its offspring's reproductive periods. Thus in adapting BP to the academic ecosystem, it is necessary to decouple *time and generation*. (b) Though the academic genealogy data provides a natural BP within the system, it is (i) sparse and not easy to obtain, (ii) not representative of the entire community. Further, one can argue that the actual reproduction takes place through publications and not through graduations. In this case, considering an author's first year of publication (year of arrival), one can consider all his senior collaborators (who have published earlier) in that year as potential parents. Thus the parent-offspring relationship will be many-many instead of one-one in the BP process. However, one can still derive insights, by observing the evolution of motifs in seniors-newcomers bipartite graphs, and by assigning fractional offspring to each parent.

In this work, we study the population growth of academic ecosystem in five parts: arrival, lifetime, activity, productivity and reproduction. While arrival, lifetime and reproduction directly influence the number of authors in the system, activity and productivity of authors also play a critical role, because reproduction occurs through publications. We use the data from the Computer Science field of MAS dataset (*22*), considering authors with at least one publication between 1960 and 2009 and all the publications written by those authors. In total, our data consists of 1,327,115 distinct authors and 2,594,871 corresponding papers. Our key findings are as follows: (i) The sustained exponential growth in the number of authors can be significantly explained by the mainstream offspring production, since the fraction of merging/immigrant authors is steadily declining (from about 91% to 22%). (ii) Motif analysis of senior-newcomer bipartite graphs reveals increasing bias towards the number of seniors than newcomers in a paper. (iii) A consistently high infant mortality rate (IMR) is observed, where about 50% authors exit after their first year of publication. (iv) Once past the first year, authors' academic lifetime tends to exhibit a geometric distribution (with $\beta = 0.95$), thus not reflecting any definitive impacts of various stages (graduate school, post-doctoral, pre-tenure and post-tenure). (v) The yearly activity rates of authors, which were around 20% until 1980, have steadily increased to about 51% in 2005, and authors also show tendency to be consecutively active in recent years. (vi) Though the individual claimed productivity (papers claimed by each author) is increasing (from 1.4 to 2.1), the community productivity (total number of papers/ total number of authors) is declining (from 1.1 to 0.7). (vii) Analysis of fractional offspring distribution reveals that the offspring likelihood steadily increases with time and the researcher's academic age.

**Arrival:**

We consider the first year of publication of an author $i$, as his year of arrival, $y_i$. We then focus on the set of *newcomers* in each year. We regard all authors with their year of arrival earlier than $y_i$ as *seniors*. For a newcomer, the papers in the first year could be (a) single-authored (b) co-authored with newcomers only or (c) co-authored with at least one senior. We refer to newcomers in (a) and (b) as *immigrants* while the newcomers in (c) are referred to as *mainstream offspring*. We observe that while both the number of mainstream offspring and immigrants are steadily increasing (Fig. 1A), the fraction of immigrants to the total number of newcomers is declining (Fig. 1B) (from 91% in 1960 to about 22% in 2009). The high early number is due to the fact that, Computer Science as a field emerged in the 1960s (*23*), and thus most authors in the early phase were technically immigrants. Another possible factor that may cause the decrease of the fraction of immigrants is that, the tendency to publish from outside the mainstream has declined (but not completely vanished), due to the maturation of the topics in computer science, which has raised the entry barrier for publication by outsiders. Thus, newcomers tend to associate themselves to existing seniors in the mainstream.

**Newcomer motifs:** Observing the collaboration patterns of newcomers in the first year (Fig. 1D) we see that immigrants tend to have fewer coauthors in their first year, than all newcomers. We also notice that since 1985, newcomers have fewer coauthors than seniors (Fig. 1C). Thus, while the newcomers have increasing number of coauthors each year, the seniors are able to leverage their wider academic network by collaborating with other seniors. Further in Fig. 1E, we see that newcomers are increasingly inclined to collaborate with seniors than other newcomers, since this helps them establish themselves in the network.

We analyze the motifs of papers involving newcomers, by considering the number of seniors (denoted by a) and other newcomers (denoted by b) involved in the paper. The motifs can be classified into six types: i) a=b=0, (single newcomer author); ii) b>a=0 (with other newcomers only) iii) b>a>0, (with more newcomers than seniors) iv) a ≥ b>0 (with no fewer seniors than newcomers) v) a=1, b=0, (with one senior only) vi) a>1, b=0, (with multiple seniors only). Among the six types of collaboration motifs, type i) and ii) represent the behavior of immigrants, and the other types represent the behavior of mainstream offspring.

Figure 1F shows the fraction of papers that fall into each motif type. Type (i) papers have steadily declined (with a rapid decrease since 1995), indicating a decreasing tendency to publish individually as a newcomer. Type (ii) papers have decreased, but only slightly, thus showing there is still tendency to publish as a group of newcomers. This might be indicative of researchers well-established outside Computer Science (say sociology or biology) publishing for the first-time as a cross-disciplinary effort. Another interesting trend we see is the steady increase in type (v) and (vi) papers, which involve one newcomer with several seniors. Type (v) papers have steadily increased till 20% and then become stable, while type (vi) papers keep increasing. It shows that seniors help newcomers establish a wider network, while also involving them in ongoing collaborations with other seniors. Fig. 1F also corroborates the increasing dominance of teams in the production of knowledge previously observed in (*24*).

**Life expectancy:**

Authors do not stay in the system indefinitely, and retire/leave the system after some time. We refer to the duration between an author's year of first publication and year of last publication as his lifetime. It is to be noted, for each year that we distinguish between authors who are *alive* and authors who are *active*. To be precise, given a year y, active authors are those who have a paper in year $y$, while alive authors have at least a paper in year $y^{'} \geq y$. Due to the limitation of the dataset time window (1960-2009), authors who arrive around 2009 may show shorter lifetime in the dataset.

One of the concepts that is closely related to life expectancy is that of infant mortality (*25*). In our context, *infant mortality rate (IMR)* refers to the fraction of newcomers who do not have papers after their first year, i.e., with lifetime equal to one year. We see that IMR is stable around 53% over the past five decades (Fig. 2A), indicating that each year, about half of the newcomers do not publish beyond their first year. A higher IMR (58.2%) has also been observed in the Scopus database (*26*). This could be explained due to several factors: (a) the newcomers (mainstream offspring) exit the system due to stringent quality filtering, academic pressure or other lucrative careers (b) for our computer science dataset, the newcomers are engineers or researchers from other domains who only publish once in the Computer Science field.

Beyond the first year, we study the life expectancy of authors, by studying the likelihood of remaining in, rather than departure from the system given the researchers' academic age. Fig. 2B shows that this is fairly independent of the academic age, and is close to 0.95. We can thus model the academic lifetime of the author, conditioned he survives beyond the first year as a geometric distribution with parameter 0.95, which indicates the average life expectancy of about 20 years. Studies on characterizing researcher's academic lifetime have shown that the "rich-get-richer" phenomenon (*27*) and contract length (*28*) are possible influencing factors, which lead them to model it using a truncated power law and an exponential tail. We also depict the CCDF (complementary cumulative distribution function) of the academic lifetime for authors arriving in different years (Fig. 2C), and observe they are similar except for boundary effects.

**Activity:**

A researcher is considered to be active if he has a publication in a given year. With increasing number of researchers and limited resources (funding, academic positions) to compete for, we would expect it to have a qualitative impact on the activity level of a researcher in recent times. However, due to the existing framework of academia with its tenure system, we might also wish to see the effect of a researcher's academic age on the activity level. While lifetime captures how long the researchers survive in the system, activity reflects on their publication behavior during the stay.

We examine the fraction of active authors to alive authors for each year, and observe that while it is relative stable and low (20%) until 1980, in recent years it has rapidly increased to about 51% in 2005. Estimates for 2009 are as high as 70% (Fig. 3A). This could be due to (a) a rapid increase in the number of publication venues (journals, conferences, workshops) in recent years (b) more visibility of peer-researchs thus revealing more open problems of interest or (c) fiercer competition for academic positions and in general, competition for attention from the community. As for the correlation between activity rate and academic age, we did not find obvious patterns in our dataset.

**Consecutive activity:** While activity in each year by itself reveals some interesting trends, one could probe further, by investigating consecutive activity of authors. For instance, we define the $CAP_N$ (*consecutively active probability for Nth year*) as the probability that an author is active at year $t + N$, given he has been consecutively active from $t$ to $t + N - 1$. We also restrict ourselves to authors who are still alive at $t + N$ (they do have a publication at or later than $t + N$). From Fig. 3B, we see that for the same t, $CAP_N$ is higher for larger $N$. For $N$ large enough ($N \geq 10$), $CAP_N$ is fairly stable around 0.95. This observation is consistent with the "rich-get-richer" phenomenon in academic publishing system (*27*), which allows prolific authors to continue to publish papers more easily as their career advances. Further, for fixed $N$, larger $t$ (closer to recent times) leads to higher $CAP_N$. This indicates researchers tend to be more consecutively active in recent times. A similar study on productivity (*26*) has shown that only about 1% of authors have uninterrupted productivity in the time window [1996-2011], and they function as the publishing core in the scientific workforce, gathering a significant fraction of the citations.

## Productivity:

It is to be noted that activity level is a better characterization of a researcher's career than lifetime, yet it is still a crude measure, since it does not take into account the number of publications in each active year. Thus we look at the productivity of the authors across time. Previous studies have shown that for individual researchers, factors like career status (*29*), career movements (*30*), aging (*31, 32*), collaboration and team efficiency (*28, 33*) may affect their productivity. Here we focus on the aggregate patterns of the entire system. For this, we define two notions: *individual productivity* (IP) is the annual number of claimed papers per author, while *community productivity* (CP) is the total number of papers divided by total number of active authors for a given year. It can also be shown that, if the number of claimed papers is balanced (based on the number of co-authors in each paper), then it reduces to CP. Fig. 4A shows that while IP is gradually increasing, CP is slightly decreasing, which is consistent with the observation in (*34*). This is to be expected, since collaborations are becoming more and more common among researchers (*24*), as shown in Fig. 1C and Fig. 1D. We also see that on restricting to a subset of authors based on their career stage, we see that CP curve is higher or lower than the cumulative CP curve (Fig. 4B). A similar effect is observed for IP (Fig. 4C). This indicates that career authors lead the productivity of the system (*6, 26*). We are also interested to find the change of IP of newcomers across different years. In Fig. 4D, we see that on average, a newcomer publishes about one paper, and this has remained fairly constant over the years. Combined with the motif analysis, it indicates that except for the collaboration patterns, newcomers' publishing behavior has not changed significantly over the years.

However, decrease in CP indicates more authors and proportionally fewer publications. This raises the question of whether the decrease in CP is due to (a) limitation in publication venues (b) increasing difficulty of research topics (c) decreased efficiency of the individual researcher. While (a) can be handled by suitably increasing the publication venues, (b) might be inherent to a mature field of research. It is to be noted that (b) could also be due to the fact that some projects require huge teams of researchers, but we suspect it is more likely in highly experimental domains like high-energy physics (*24*) and not very prevalent in Computer Science.

More importantly, (c) seems to be troubling, since it could either mean that authors become less efficient due to switching between several projects, or since authors tend to be evaluated based on IP (*35, 36*) which does not take into account the number of coauthors in each paper (which is indeed increasing according to Fig. 4E). If this is indeed the case, then one could investigate better collaboration mechanisms, and better ways to evaluate the productivity of an individual researcher (*28*).

**Reproduction:**

As noted in Fig. 1B, the exponential growth in community size can be explained significantly by the production of mainstream offspring. Understanding the reproduction behavior, not only provides insights for the observation of the rapid growth of authors and publications, but can also be used to study correlation between fertility across different generations (*21, 37*). Borrowing from the branching process terminology, we use $p_{s,t}(k)$ to represent the probability that a researcher arriving at time $s$, has $k$ offspring at time $t$, given that he is *alive and active* at time $t$. However, as mentioned earlier, we intend to define parent-offspring relation not based on the academic lineage (graduations), but based on the publications in the first year of an author. Thus, a newcomer could have potentially more than one parent. In order to resolve this issue, we assign fractional offspring to each senior collaborator as follows: Each newcomer's credit is shared equally among all his papers in the first year, and further for each paper, equally among all senior collaborators. The fractional offspring for a particular senior in a given year is then, the total credit he receives from all newcomers in that year. This scheme has advantages over other ways of assigning such as all-all assignment (where each senior gets full credit for every newcomer he collaborates with) or all-one assignment (where only one of the seniors gets full credit), which either do not preserve the network structure or the size. Thus, $k$ can be fractional.

To begin with we look at the case when $k = 0$, i.e., the probability that a senior researcher does not collaborate with any newcomers in a given year. We first observe that, for fixed $s$, $p(0)$ (short for $p_{s,t}(0)$) is decreasing as $t$ increases (Fig. 5A). This indicates that seniors tend to collaborate with newcomers more often, as their academic age increases. For fixed $t$, as $s$ increases, $p(0)$ remains stable until $t - s = 10$, and then is non-decreasing in $s$ (Fig. 5B). This indicates that researchers in the early-career phase (until about academic age of 10) tend to collaborate more with seniors, and later have a fairly constant and low $p(0)$.

Having observed the pattern in $p_{s,t}(0)$, we wish to observe the distribution for $p_{s,t}(k)$. Fig. 5C and Fig. 5D shows the CDF (cumulative distribution function) of $p_{s,t}(k)$ when fixed $s$ (Fig. 5C) and fixed $t$ (Fig. 5D) respectively. For fixed $s$, we can see that except for the change in

$p_{s,t}(0)$ across different $t$ s, $p_{s,t}(k)$ keeps almost consistent for $k$ greater than or equal to 1, and exhibits a light tail. To be more specific, when $k$ is equal to 1, the fraction of senior researchers who have fractional offspring no larger than 1, is around 87% over different $t$ s. A similar characteristic is observed for fixed $t$. Thus, we see that $p_{s,t}(k)$ for different $s$ and $t$ have qualitatively similar distributions and might be characterized by $p_{s,t}(0)$.

**Discussions:**

To our knowledge, though the existing publication datasets have been subjected to various network structural analysis in the past (*4, 6*), not much attention has been given to understanding the influencing factors for the exponential growth of the system. Our work is a first attempt to capture this phenomenon by building a population model for the academic community. Such a model creates a systematic method to describe the size and composition of the community (in terms of immigrants, seniors and their offspring) as well as the size and composition over time. Our population model is inspired by previous studies of human communities of different scales, and the basic branching process model, a key theoretical foundation for previous studies. Our study, however, is not a trivial application of the branching process model, but uses it as a guiding framework for analyzing various factors within the system. The model can be further refined by including resource (funding, faculty positions) constraints.

Based on the framework created by the population model, and the publication data we have for the computer science community, we are able to uncover many interesting properties. Some appear to be invariant over time (infant mortality rate, life time distribution), some changes slowly over time (offspring distribution, productivity), and some others are changing significantly over time (ratio of immigrants, activity levels). These insights demonstrate the power and the need for such a population model. The analysis was also repeated on the DBLP dataset (*38*) for the computer science community (see the Supplementary Material) in order to avoid the bias of the MAS dataset. Similar conclusions were obtained, which shows the robustness of our findings. Results from our work could help improve scientific evaluation methods by accounting for inflation, and in general, aid in policy-making for better resource allocation in academia.

**Acknowledgments:** The MAS dataset is collected from the Microsoft Academic Search public API. The DBLP dataset is publicly available on the website.

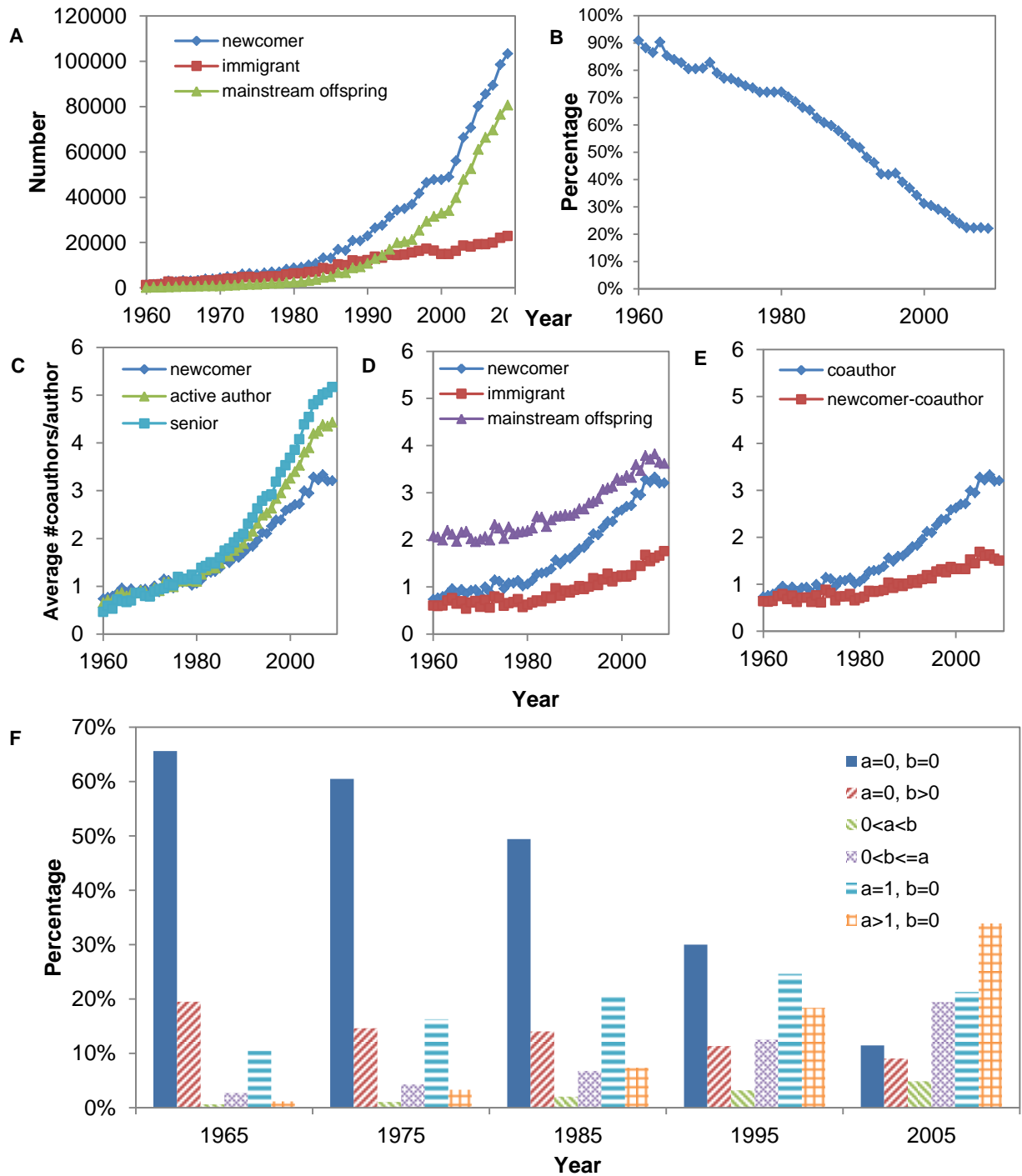

**Fig. 1**. **Arrival.** (**A**) Number of newcomers, immigrants, and mainstream offspring each year between 1960 and 2009. (**B**) Fraction of immigrants among the total number of newcomers each year. (**C**) Average annual number of coauthors for active authors, newcomers and senior researchers. (**D**) Average annual number of coauthors for newcomers, immigrants and mainstream offspring. (**E**) Average annual number of coauthors and newcomer-coauthors for newcomers. (**F**) Fraction of papers belonging to each motif type in different years.

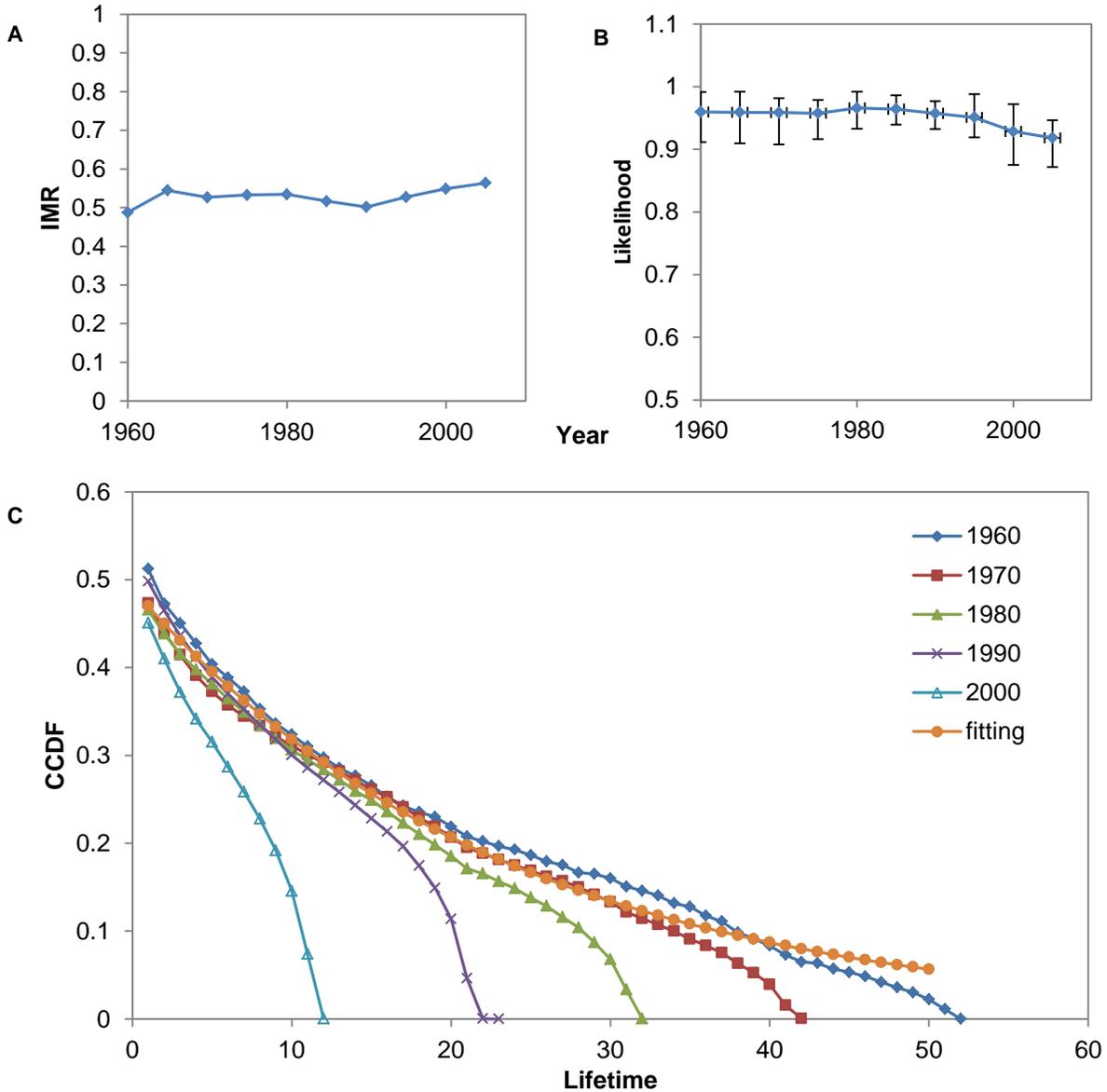

**Fig. 2. Life expectancy. (A)** The value of IMR every year. **(B)** The likelihood of remaining in, rather than departure from the system given the researcher's academic age every year. The marker shows the average and the error bar shows the range among researchers with different age at a specific year. **(C)** Academic lifetime distribution (CCDF) of authors arriving at different years in the dataset and a geometric distribution fitting curve with IMR=0.5095 and $\beta$ =0.9577.

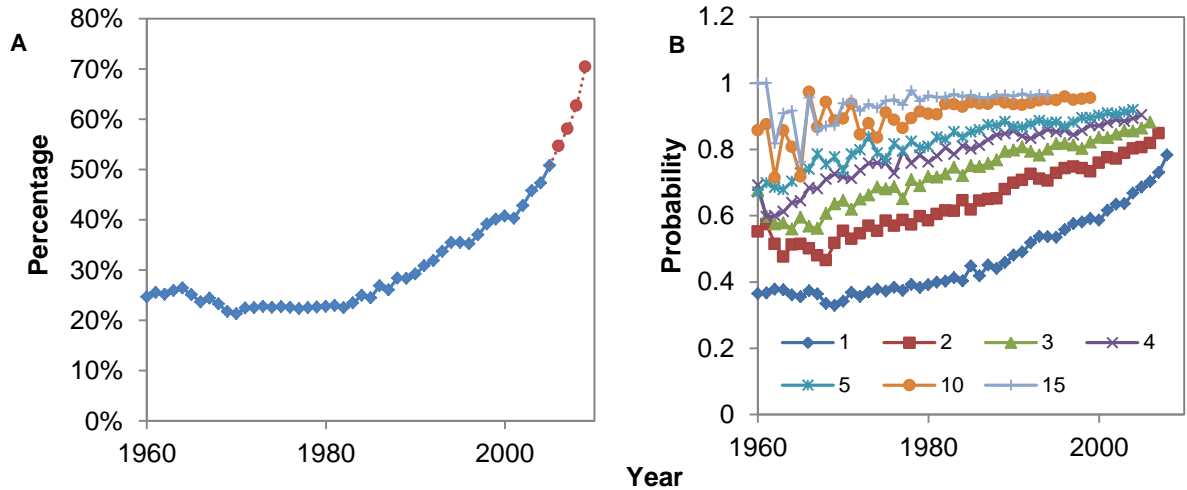

**Fig. 3. Activity.** (**A**) Fraction of active authors among alive authors every year, indicating the active extent in the academic ecosystem. Due to the boundary effect of the dataset, alive authors in recent years may be underestimated since they might stay inactive rather than exit. In our dataset, about 26% of the authors have inactive period during their career, among which 75% have longest inactive period no larger than five years, thus we set a five-year window on the longest inactive period. (**B**) Change of $CAP_N$ at different time $t$ for different values of $N$.

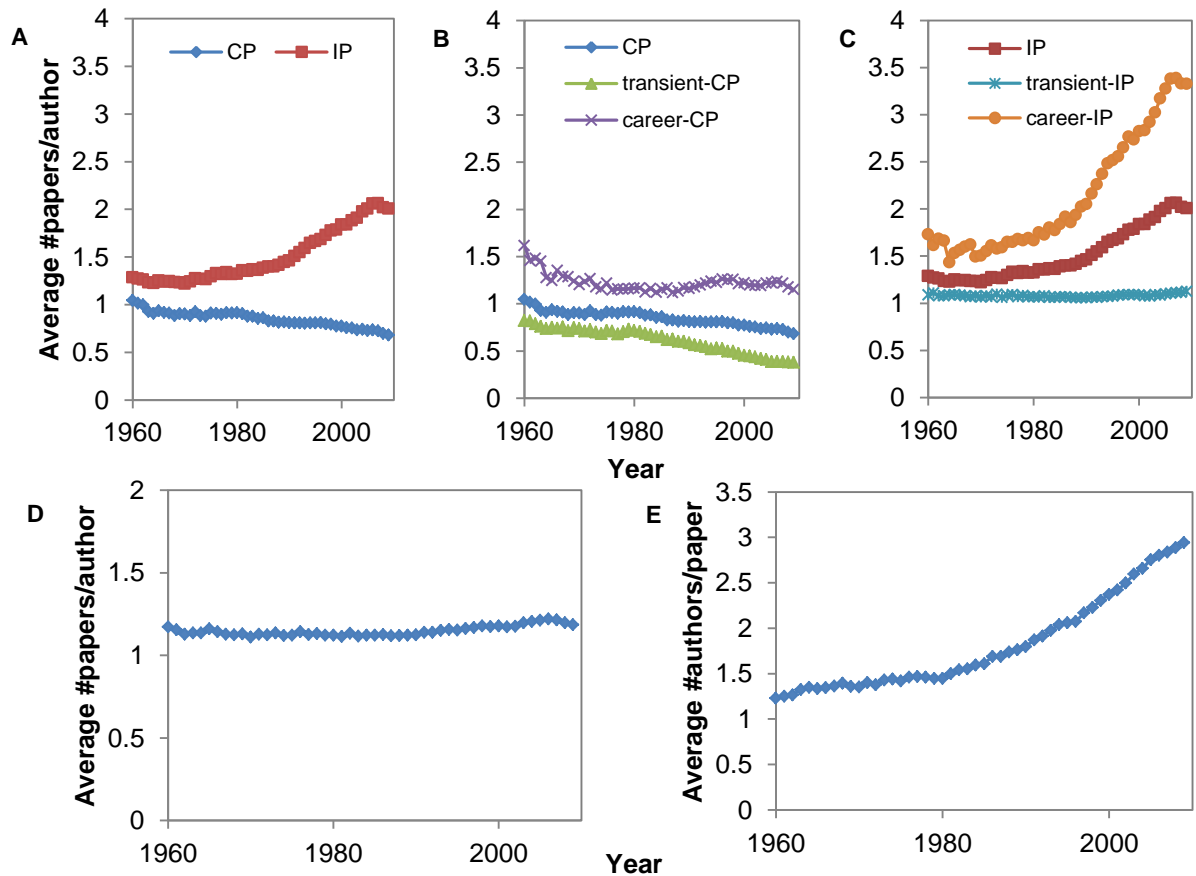

**Fig. 4. Productivity. (A)** Change of CP and IP for active authors. **(B)** Comparison of CP for transient authors, career authors and active authors. Transient authors are defined as authors who only appear for one year in our dataset. Career authors are defined as authors whose current academic age is not smaller than ten at time $t$. **(C)** Comparison of IP for transient authors, career authors and active authors. **(D)** Change of IP for newcomers at different years. **(E)** Average number of authors in each paper every year.

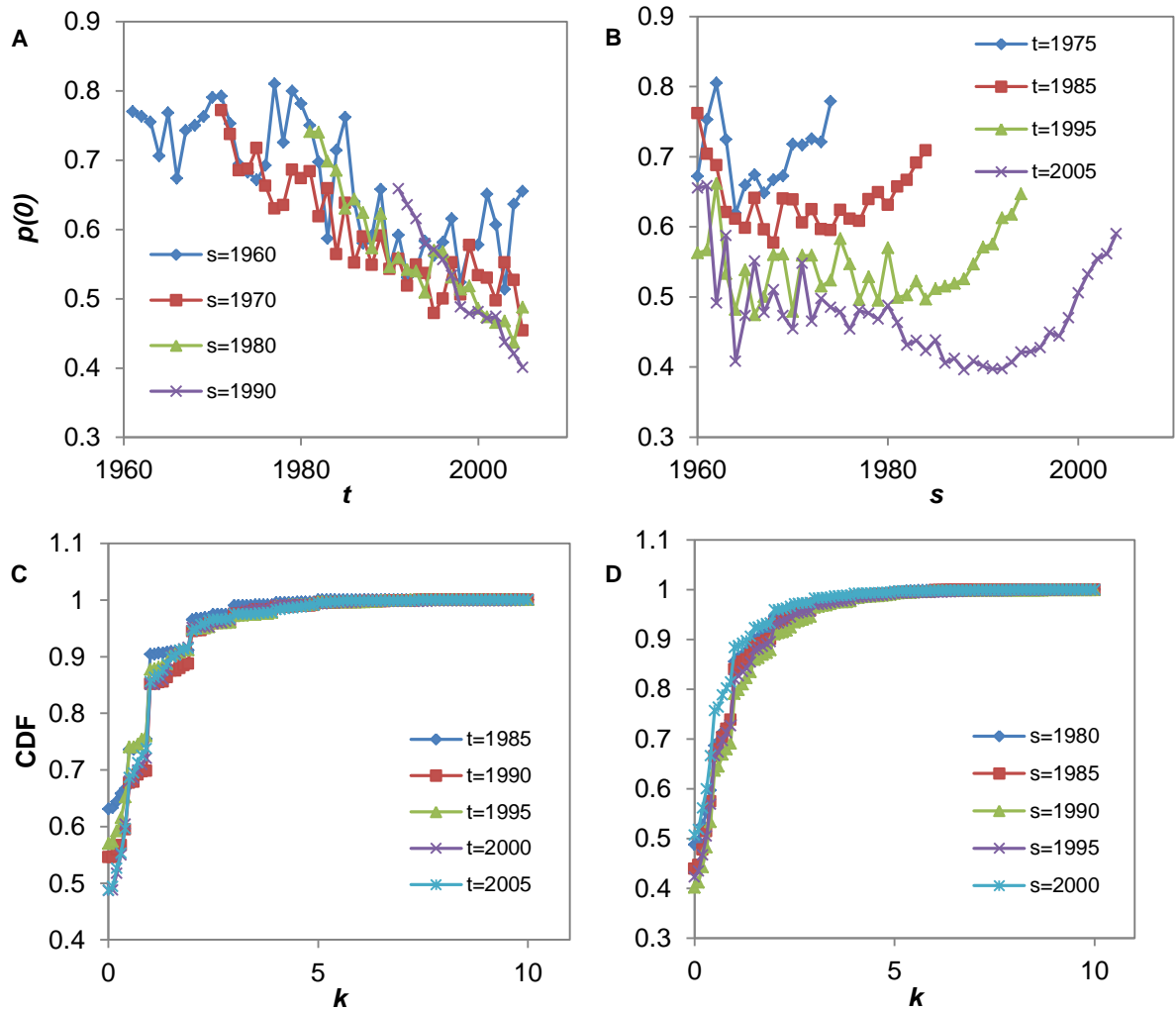

**Fig. 5. Reproduction. (A)** Change of $p(0)$ at different time t when fixed the senior researcher's arrival time $s$. **(B)** Change of $p(0)$ for senior researchers arriving at different time $s$ when fixed current time $t$. **(C)** CDF of $p_{s,t}(k)$ at different time t when fixed $s = 1980$. **(D)** CDF of $p_{s,t}(k)$ at different time $s$ when fixed $t = 2005$.

**Supplementary Materials:**

Supplementary Text

Figures S1-S8

Tables S1

**Supplementary Materials:**

**MAS Dataset:**

Our data is collected through the Microsoft Academic Search (MAS) public API (*22*). The MAS publication database is classified into a wide range of (15) research fields, which are further categorized into domains within that field. For example, in Computer Science, it includes 24 domains such as "Algorithms and Theory", "Security and Privacy", "Hardware and Architecture", etc. Each paper is assigned a unique numerical ID and the database includes information such as its title, author list, publication year, venue of publication and the list of references. Further, each author is assigned a unique numerical ID, with information including current affiliation and publication history. We use this author ID as is, and do not perform further disambiguation. Other information like the author's research field can be obtained from his publication history. We choose the Computer Science field which we are most familiar with as a case study in this paper, and we believe our methodology can be replicated for other research fields as well.

Due to incomplete electronic records for the early years, we take the data in year range [1960, 2009] for analysis in the paper, spanning 50 years. This time window is sufficient to capture the evolution of Computer Science as a discipline, since its origins are traced back to early 1960s (*23*).

**Supplementary figures in MAS dataset:**

Figure S1 shows the rapid increase in the number of alive authors, active authors and publications each year. Due to the boundary effect, the number of alive authors may be underestimated in recent years. From Fig. S1 we see that the rate of increase of authors is even higher than that of publications.

Figure S2 shows the percentage of newcomers among active authors each year. Note that most of the early authors are newcomers by definition, since the corresponding papers are among the earliest classified under Computer Science. Later, although the absolute number of newcomers continues to increase, the percentage has declined to about 40% in recent years, indicating that the population has more senior authors than newcomers. Figure S3 shows a similar phenomenon, for the number of papers involving at least one newcomer each year.

**Branching Process Model:**

We provide the mathematical formulation of our population model based on the generalized branching process, starting with the introduction of notations. Let $N(t)$, $B(t)$ and $D(t)$ be the number of alive authors, newcomers and authors exiting the system at time $t$ respectively. Then

$$N(t) = N(t-1) + B(t) - D(t) . \tag{S1}$$

The parameters required to describe each term are life expectancy, activity and reproduction of each newcomer. Let $q(l)$ to be the probability that an author has academic lifetime $l$; $\theta(t)$ to be the probability that an author is active at time $t$, given that he is still alive at time $t$; and $p_{s,t}(k)$ to be the probability that an author arriving at time $s$ has fractional offspring $k$ at time $t$, given that he is alive and active at time $t$. Note that for newcomers, they are alive and active in their year of arrival, by definition.

Then we have

$$N(t) = N(0) + \sum_{s=1}^{t} [B(s) - D(s)], \tag{S2}$$

$B(t)$ consists of two parts, namely, the immigrants and mainstream offspring:
$B(t) = I(t) + M(t)$. We use $I(t)$ to represent the expected number of immigrants at time $t$ and focus on the change of mainstream offspring $M(t)$. $M(t)$ can be represented as

$$M(t) = \sum_{s<t} \{B(s)[1 - \sum_{j=s+1}^{t} q(j-s)]\theta(t) \sum_{k>0} k p_{s,t}(k)\}, \tag{S3}$$

which is the mainstream offspring generated by seniors who are still alive and active at time $t$. Note that $k$ is not restricted to integers, and can take rational values.

$D(t)$ is determined by the academic lifetime distribution $q(l)$.

$$D(t) = \sum_{s<t} B(s) q(t-s). \tag{S4}$$

As described in the main text, $q(l)$ is characterized by IMR (we represent by the parameter $\alpha$) and the geometric parameter $\beta$, and can be represented as

for $l = 1$, $q(l) = \alpha$; \hfill (S5)

for $l > 1$, $q(l) = (1 - \alpha)\beta^{l-2}(1 - \beta)$. \hfill (S6)

Similarly, one can obtain a $\theta(t)$ from the activity analysis. We also see that $p_{s,t}(k)$ can be parameterized by $p_{s,t}(0)$ and can help characterize the expected offspring for each $s$ and $t$, given by $\mu_{s,t} = \sum_{k>0} k p_{s,t}(k)$ (see Table S1, for example).

**DBLP Dataset:**

DBLP is a bibliography project hosted by Trier University (*38*), and covers most of the important journals and proceedings of conferences in the field of Computer Science. It provides open access to its dataset via XML format which shows detailed information based on each author and each paper. For each author, it gives the publication list of that author based on publication year. For each paper, it gives the paper title, author list, publication year and publication venue. Our DBLP dataset is collected via (*39*), which extracts information from the original XML file and gives us better data structure for later analysis. In order to compare the statistics in DBLP dataset with the MAS dataset, we also restrict the DBLP dataset to the time window [1960, 2009]. In the DBLP dataset, the number of authors with at least one publication in that period is 838,185. And the number of papers written by those authors in the DBLP dataset is 1,548,785.

**Results in the DBLP Dataset:**

Figure S4-S8 show the corresponding results in the DBLP dataset, which corroborate our findings on the MAS dataset (Fig. 1-5 in the main text). Due to the incomplete coverage of the early years in the DBLP database, they exhibit high fluctuations for some characteristics, and we have removed the curve for year 1960 in Fig. S5C and Fig. S8A. Since each dataset is collected independently, we believe the similar findings obtained in both datasets show the robustness of our conclusions in this paper.

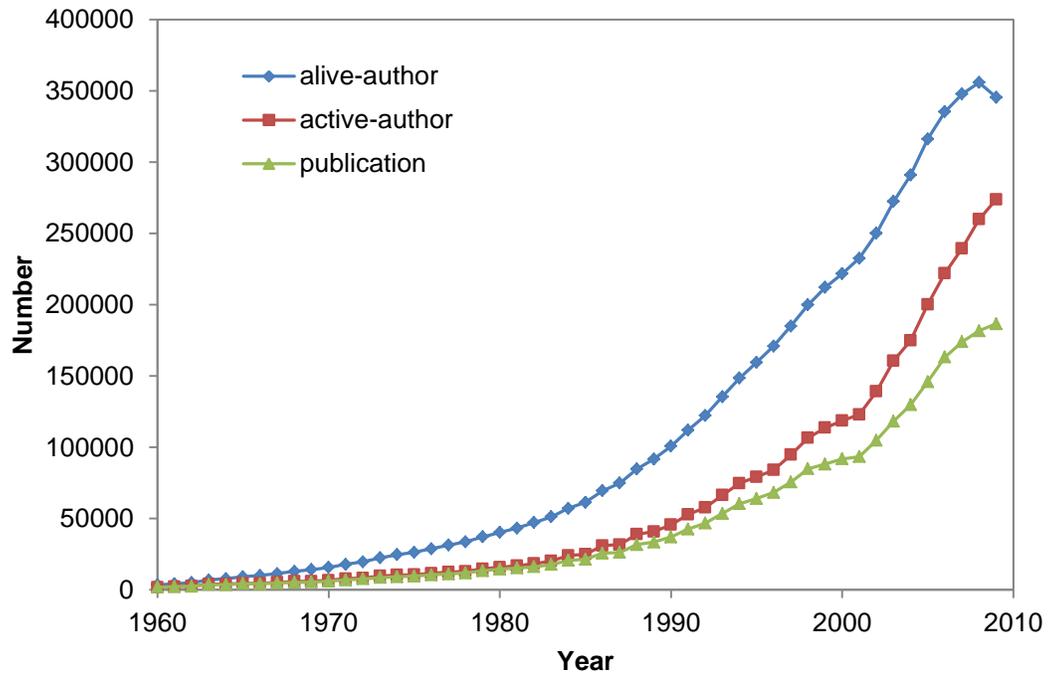

**Fig. S1.** Number of alive authors, active authors and publications each year between 1960 and 2009 (MAS dataset).

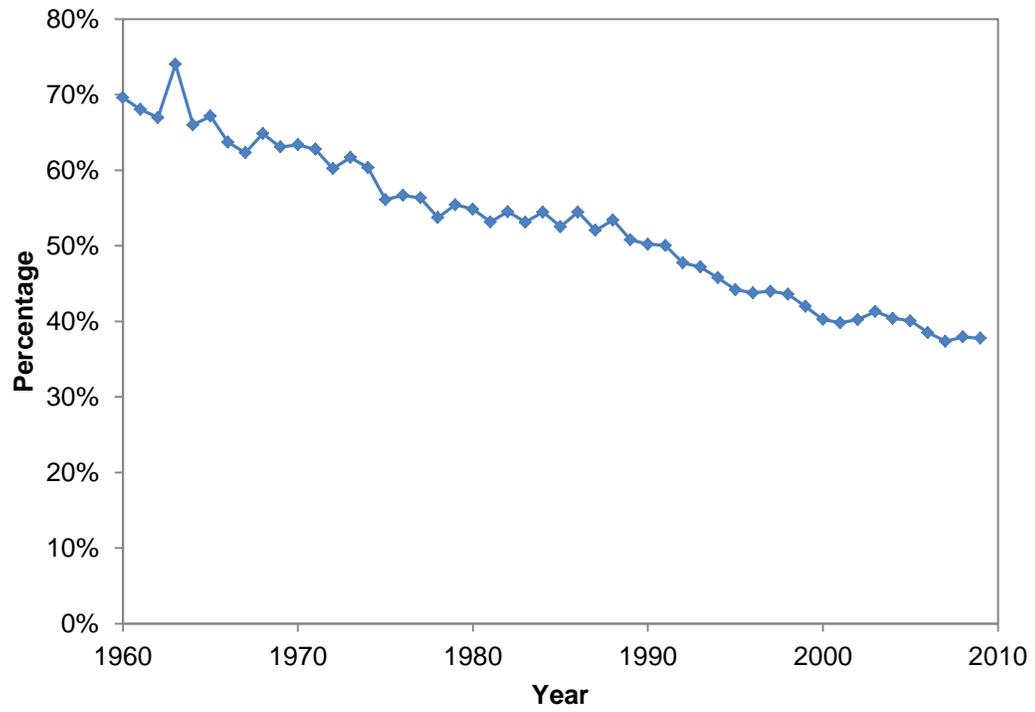

**Fig. S2.** Percentage of newcomers among active authors each year between 1960 and 2009 (MAS dataset).

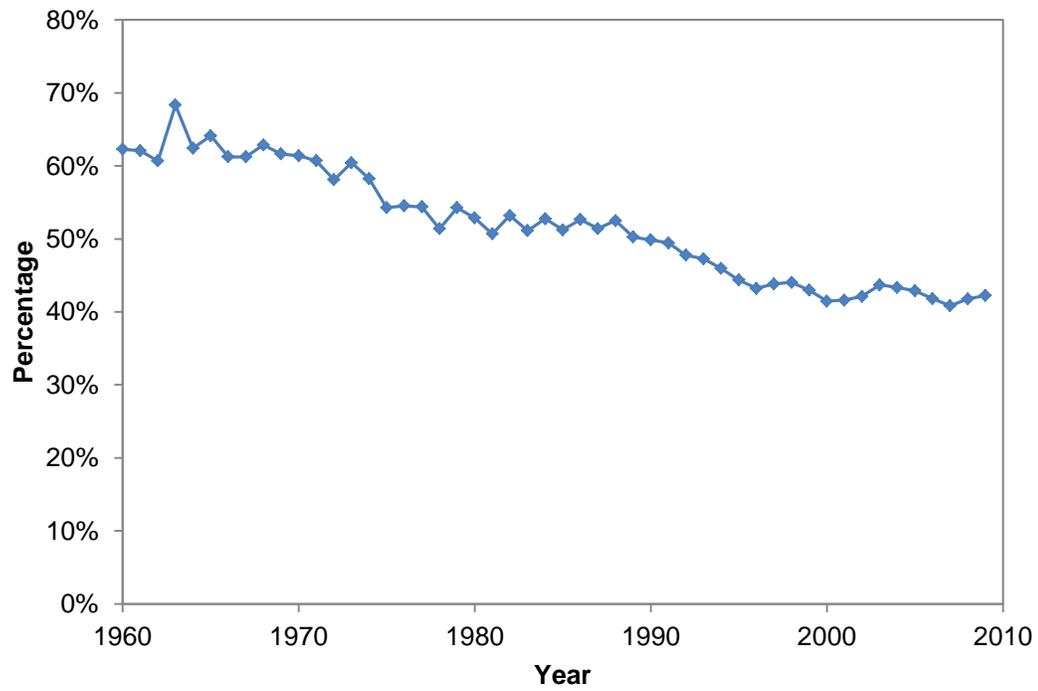

**Fig. S3.** Percentage of papers involving newcomers each year between 1960 and 2009 (MAS dataset).

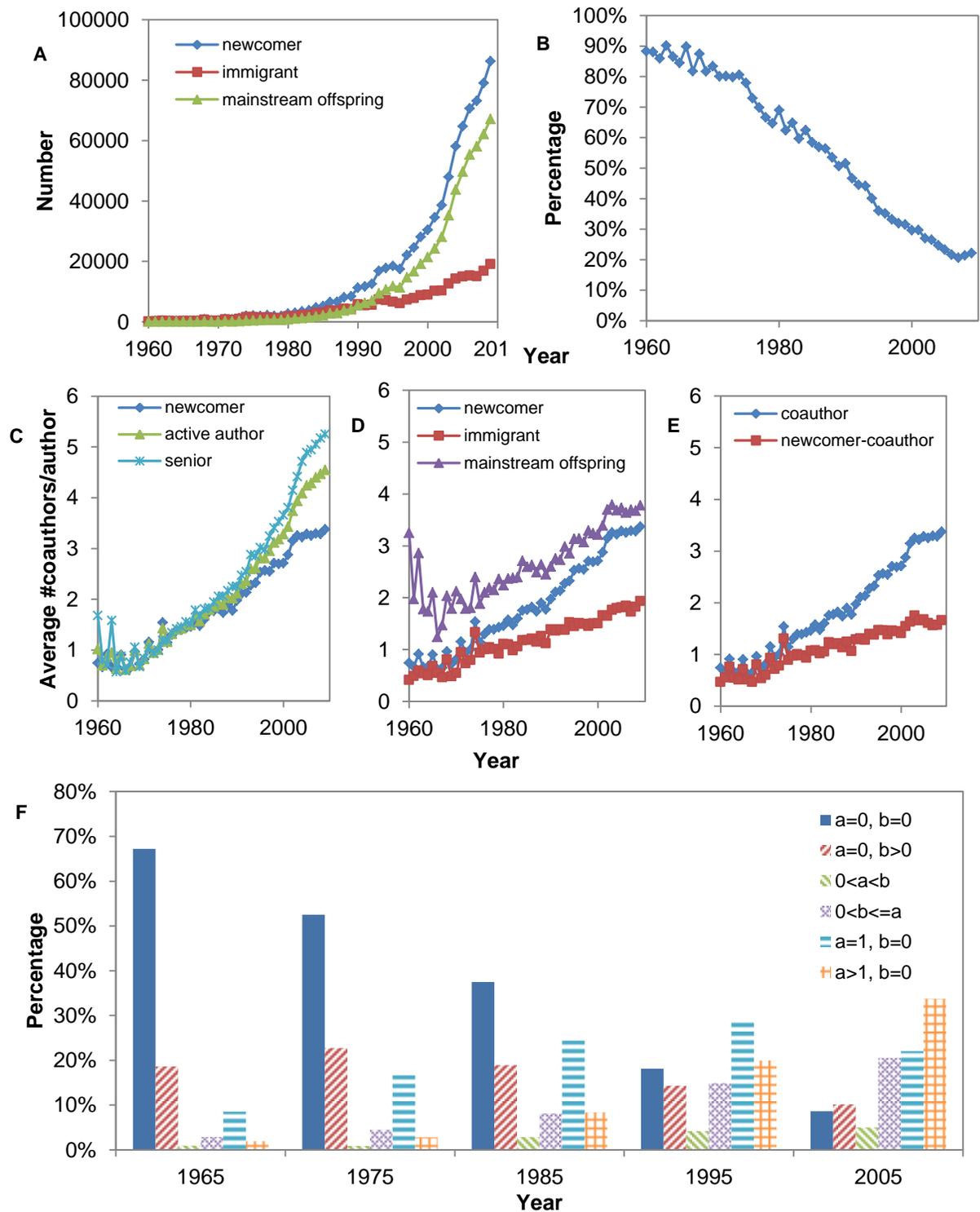

**Fig. S4. Arrival.** (**A**) Number of newcomers, immigrants, and mainstream offspring each year between 1960 and 2009. (**B**) Fraction of immigrants among the total number of newcomers each year. (**C**) Average annual number of coauthors for active authors, newcomers and senior researchers. (**D**) Average annual number of coauthors for newcomers, immigrants and mainstream offspring. (**E**) Average annual number of coauthors and newcomer-coauthors for newcomers. (**F**) Fraction of papers belonging to each motif type in different years. (DBLP)

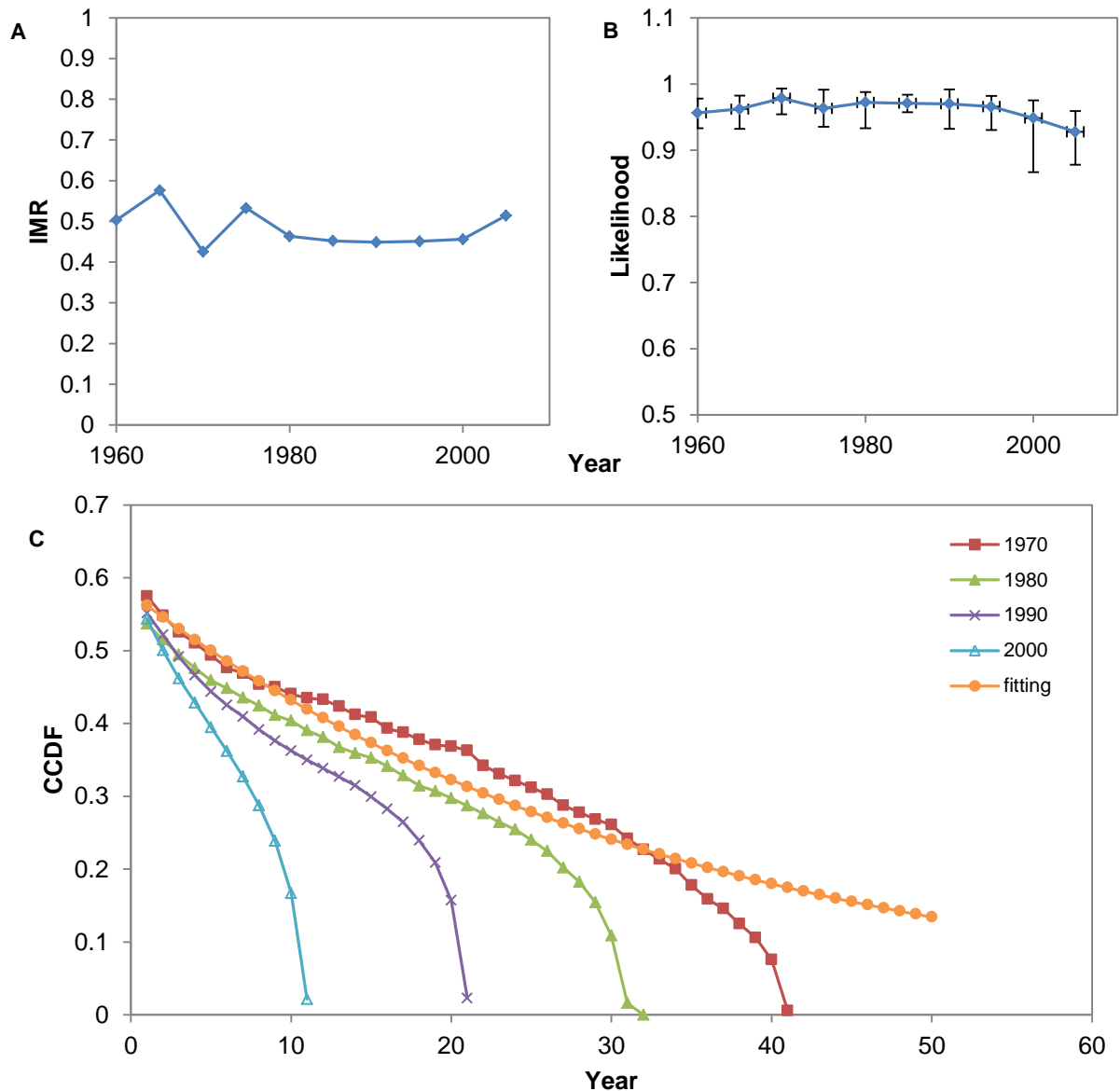

**Fig. S5. Life expectancy. (A)** The value of IMR every year. **(B)** The likelihood of remaining in, rather than departure from the system given the researcher's academic age every year. The marker shows the average and the error bar shows the range among researchers with different age at a specific year. **(C)** Academic lifetime distribution (CCDF) of authors arriving at different years in the dataset and a geometric distribution fitting curve with IMR=0.4214 and $\beta$ =0.9712. (DBLP)

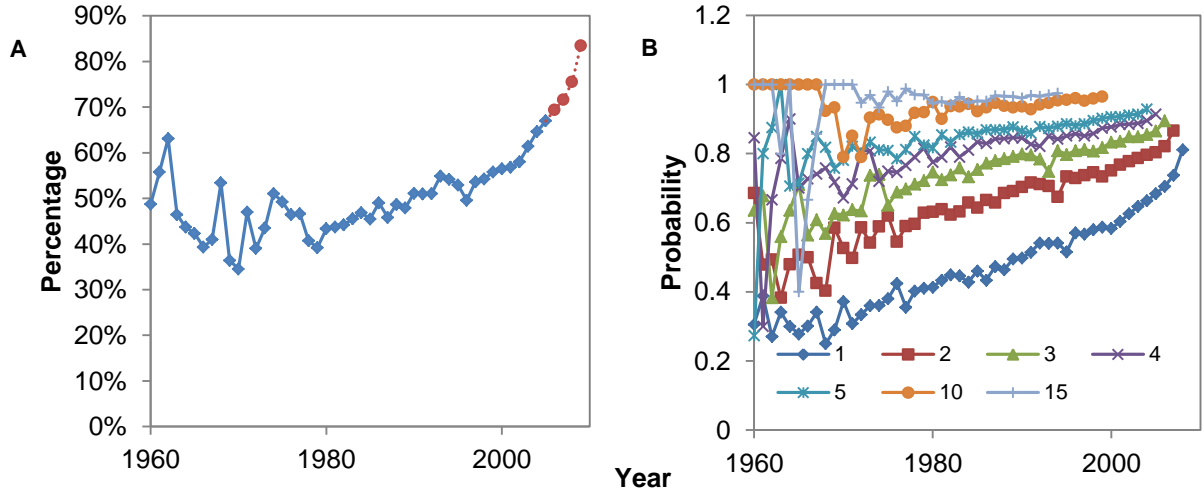

**Fig. S6. Activity. (A)** Fraction of active authors among alive authors every year, indicating the active extent in the academic ecosystem. Due to the boundary effect of the dataset, alive authors in recent years may be underestimated since they might stay inactive rather than exit. So we set a five-year window on the longest inactive period. **(B)** Change of $CAP_N$ at different time $t$ for different values of $N$. (DBLP)

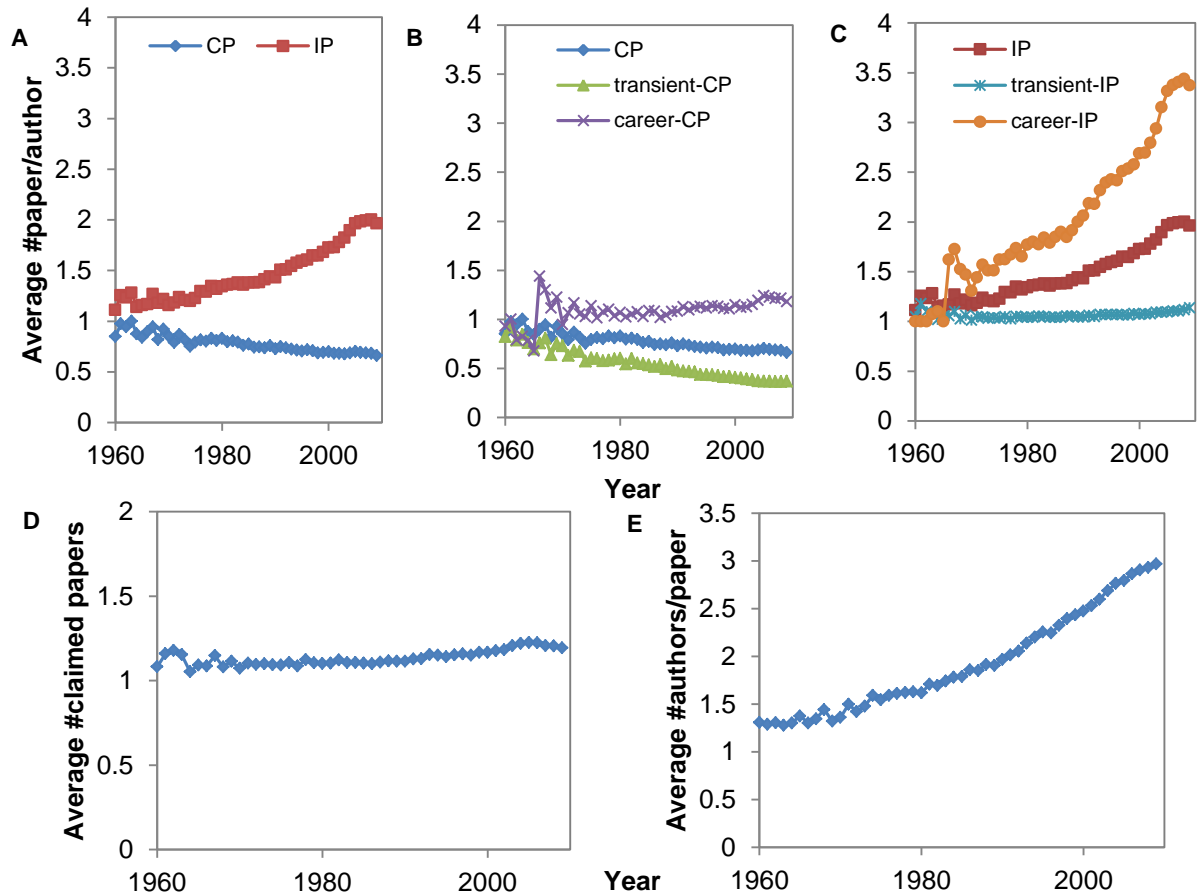

**Fig. S7. Productivity. (A)** Change of CP and IP for active authors. **(B)** Comparison of CP for transient authors, career authors and active authors. Transient authors are defined as authors who only appear for one year in our dataset. Career authors are defined as authors whose current academic age is not smaller than ten at time $t$. **(C)** Comparison of IP for transient authors, career authors and active authors. **(D)** Change of IP for newcomers at different years. **(E)** Average number of authors in each paper each year. (DBLP)

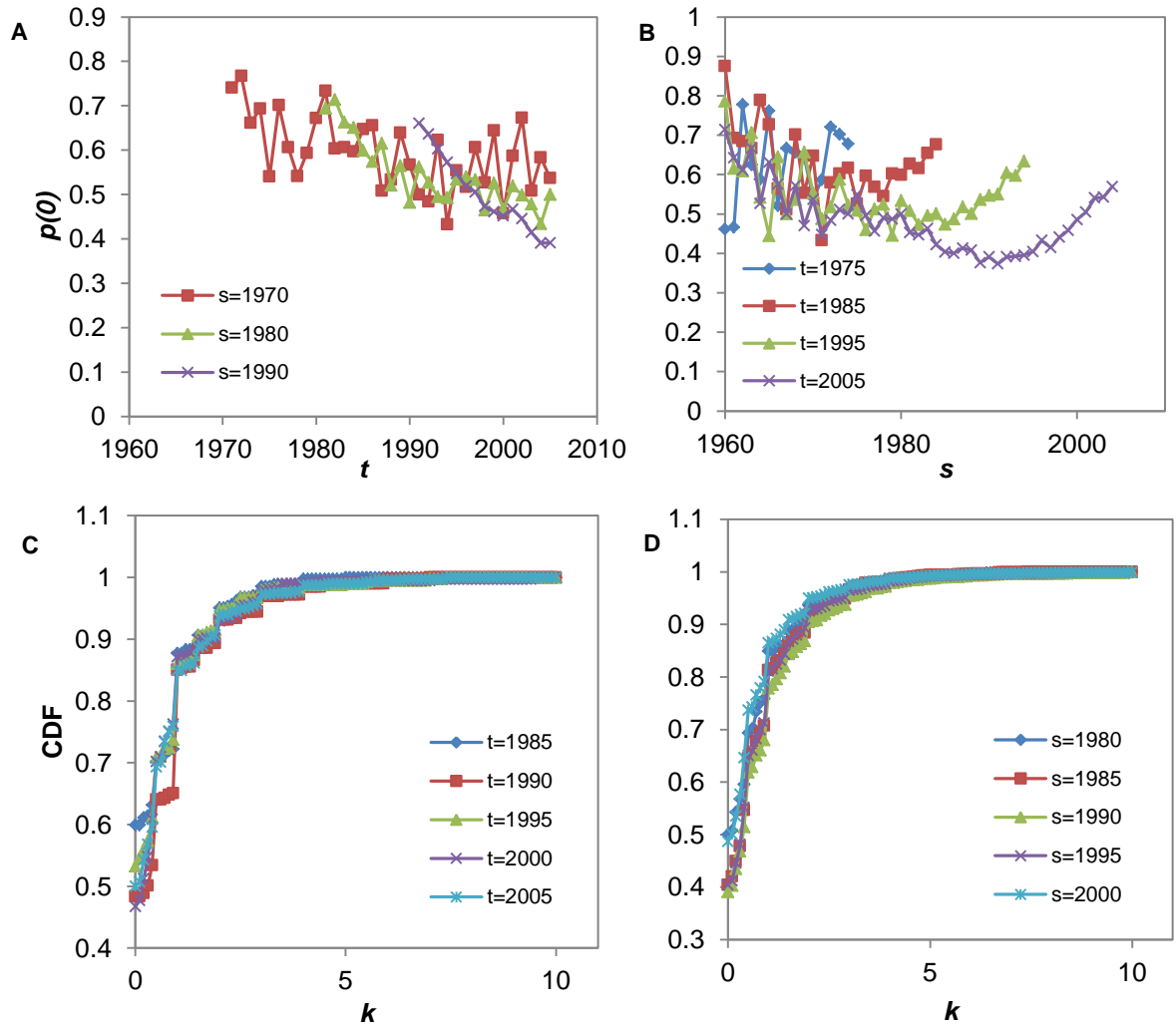

**Fig. S8. Reproduction.** (**A**) Change of $p(0)$ at different time $t$ when fixed the senior researcher's arrival time $s$. (**B**) Change of $p(0)$ for senior researchers arriving at different time $s$ when fixed current time $t$. (**C**) CDF of $p_{s,t}(k)$ at different time $s$ when fixed $s=1980$. (**D**) CDF of $p_{s,t}(k)$ at different time $s$ when fixed $t=2005$. (DBLP)

**Table S1.** $\mu_{s,t}$. **(A)** Change of $\mu_{s,t}$ at different time $t$ when fixed $s = 1980$. **(B)** Change of $\mu_{s,t}$ at different time $s$ when fixed $t = 2005$. (MAS dataset)

A

| $t$ | 1985 | 1990 | 1995 | 2000 | 2005 |
|---|---|---|---|---|---|
| $\mu_{s,t}$ | 0.4359255 | 0.5850305 | 0.5090113 | 0.5735119 | 0.5596963 |

B

| $s$ | 1980 | 1985 | 1990 | 1995 | 2000 |
|---|---|---|---|---|---|
| $\mu_{s,t}$ | 0.5596963 | 0.6152941 | 0.7157645 | 0.646298 | 0.4589691 |